\documentclass[bibyear]{aa}
\usepackage{amsmath,amstext}
\usepackage{natbib,twoopt}
\usepackage[breaklinks=true,colorlinks=true,linkcolor=blue,citecolor=blue,urlcolor=blue]{hyperref}
\usepackage{wrapfig}
\usepackage{amssymb}
\usepackage{wasysym}
\usepackage{float}
\usepackage{ulem}
\usepackage{breakurl}
\usepackage{color}
\usepackage[flushleft]{threeparttable} 
\newcommand{\red}{\color{black}}

\def\spose#1{\hbox to 0pt{#1\hss}}
\def\lta{\mathrel{\spose{\lower 3pt\hbox{$\mathchar"218$}}
     \raise 2.0pt\hbox{$\mathchar"13C$}}}
\def\gta{\mathrel{\spose{\lower 3pt\hbox{$\mathchar"218$}}
     \raise 2.0pt\hbox{$\mathchar"13E$}}}

\begin{document}

\newenvironment{tablehere}
  {\def\@@captype{table}}
  {}
\newenvironment{figurehere}
  {\def\@@captype{figure}}
  {}
\makeatother
\makeatletter

\titlerunning{Obscured little blue dots}

\title{Little red dots as obscured little blue dots: Relative abundances, luminosities, and black-hole masses}

\author{Piero Madau\inst{1,2}
\and
Roberto Maiolino\inst{3,4,5}
}
\institute{Dipartimento di Fisica ``G. Occhialini,'' Università degli Studi di Milano-Bicocca, Piazza della Scienza 3, I-20126 Milano, Italy \and Department of Astronomy \& Astrophysics, University of California, 1156 High Street, Santa Cruz, CA 95064, USA \and
Kavli Institute for Cosmology, University of Cambridge, Madingley Road, Cambridge CB3 0HA, UK 
\and Cavendish Laboratory, University of Cambridge, 19 JJ Thomson Avenue, Cambridge CB3 0HE, UK \and
Department of Physics and Astronomy, University College London, Gower Street, London WC1E 6BT, UK}

\abstract{We test whether ``little red dots'' (LRDs) are the dust-reddened, high-inclination counterparts of bluer compact broad-line active galactic nuclei, here referred to as ``little blue dots'' (LBDs), by modeling their relative number densities and luminosities. Using the observed UV luminosity function (LF) of broad-line active galactic nuclei (BLAGNs) at $z\gtrsim4$ as the parent distribution, we forward-model the effects of accretion rate, anisotropic emission, orientation, and dust obscuration within our super-Eddington unification framework. We show that a model with a geometrically thick accretion flow and a dusty circumnuclear cloud population reproduces the LRD LF over the luminosity range currently constrained by JWST. The predicted LRD/BLAGN fraction is strongly luminosity dependent, rising from $\simeq 3\%$ at $M_{1500}=-21$ mag to a peak value of $\sim 20\%$ near $M_{1500}\simeq -19$. The model also predicts a larger apparent LRD fraction at rest-frame optical wavelengths, reaching 27\% at $M_{4500}=-20$ mag and 39\% at $M_{6500}=-21$. {\red The fiducial solution adopts a baseline host extinction $A_V^{\rm host}=0.5$~mag, a characteristic per-cloud nuclear extinction $\langle A_V^c\rangle=3.0$~mag, and a gray $R_V=5.5$ attenuation curve, and yields a mean dust covering factor $\langle C_{\rm dust}\rangle\simeq0.23$.} These results may support an orientation-based unification of little dots and identify the LRD LF as a key demographic test of rapid accretion onto infant black holes at cosmic dawn. {\red In the AGN-dominated model, LRDs and LBDs are drawn from the same observed BLAGN UV LF, but attenuation makes UV-selected LRDs intrinsically brighter and their inferred black holes more massive at fixed observed $M_{1500}$.}
}


\maketitle
\section{Introduction}
\label{sec:intro}

The James Webb Space Telescope has uncovered a large population of \textcolor{black}{BLAGNs} at $z \gtrsim 4$, powered by accretion onto massive black holes with inferred masses of $\sim10^{6}$--$10^{8}\,M_\odot$
\citep[e.g.,][]{Harikane2023AGN,MaiolinoAGN,Taylor2025_BHMF,Juod2026}. Within this emerging class, a subset of sources exhibits red, V-shaped UV-optical continua and is commonly classified as LRDs, while the remainder show bluer continua and are often referred to as LBDs \citep{Brazzini2026}. Recent analyses suggest that these categories may not represent two sharply
distinct populations, but rather the opposite ends of a continuous sequence in compactness, continuum shape, and broad-line prominence \citep{Billand2026}. In current JWST spectroscopic samples, LRDs appear to constitute only $\sim10\%$--30\% of the \textcolor{black}{BLAGN population} once selection effects are taken into account -- in particular the fact that LRDs have often been preferentially targeted because of their unusual colors, whereas LBDs, having colors similar to star-forming galaxies, are typically identified only
serendipitously in spectroscopic surveys \citep{Hainline2025,Taylor2025_BHMF}.

In Paper~I \citep{MadauMaiolino2026}, we proposed that LRDs and LBDs are different viewing-angle manifestations of the \textcolor{black}{same rapidly accreting BLAGN population}. In that framework, the central engine is a geometrically thick, radiation-supported accretion flow that emits anisotropically and is surrounded by an equatorially concentrated BLR and a dusty circumnuclear component of modest covering factor. LBDs correspond to lower-inclination, relatively unobscured sightlines, whereas LRDs are the same sources viewed through dust-intercepted lines of sight. Because both the intrinsic anisotropic continuum and the line-of-sight attenuation depend on viewing angle, the observed LRD/LBD ratio inferred from spectroscopic surveys is not a purely geometric quantity. It reflects instead the convolution of orientation-dependent luminosity, photometric selection, and dust obscuration.

The physical nature of LRDs remains debated. Purely stellar interpretations
invoke compact, evolved stellar populations with strong nebular emission to explain the V-shaped SEDs \citep[e.g.,][]{Baggen2024,Perez2024,Leung2025,Labbe2025,Hainline2025}. However, several lines of evidence increasingly favour an AGN-dominated continuum in at least a substantial fraction of LRDs: the prevalence of broad Balmer lines and extreme Balmer EWs \citep{Hviding2025,Yan2025}, the frequent
dominance of an unresolved rest-optical component \citep{Killi2024,Hviding2025},
and spectroscopic features around the Balmer break that are more naturally
explained by absorption and radiative transfer in high-density gas than by an
evolved stellar population
\citep{Juod2024JADES,Ji2025,Inayoshi2025,DEugenioI2025,Yan2025}.

A key uncertainty concerns the origin of the rest-UV continuum used to construct the LRD and BLAGN luminosity functions. It has been argued that, even if the rest-optical emission is AGN-related, the UV spectrum may still be dominated by stellar light \citep[e.g.,][]{deGraaff2025,Baggen2026}. At the same time, detailed analyses of individual LRDs and related compact broad-line systems have found that the UV continuum is often consistent with being primarily AGN-powered, albeit with some host-galaxy contribution in certain
cases \citep[e.g.,][]{Labbe2024,DEugenioBT2025,Ji2026,Tang2026,Pacucci2026,Torralba2026}. In what follows, we adopt this AGN-dominated UV hypothesis as the working assumption of our LF modeling.

In this paper we test that unification picture at the population level. We ask whether the observed LRD LF can be recovered from the \textcolor{black}{BLAGN parent population} through orientation-dependent emission and dust interception alone. Using the observed BLAGN UV LF as input, we forward-model the effects of accretion rate, viewing angle, anisotropy, and extinction on the emergent rest-UV continuum. The central question is whether the same super-Eddington geometry that links LBDs and LRDs at the level of individual SEDs can also reproduce the observed abundance and luminosity dependence of LRDs with plausible dust covering factors and extinction distributions. Failure to do so would imply that orientation and obscuration alone are insufficient, and that additional physical distinctions between the two populations are required.

\section{Hybrid forward modeling of the LRD LF}

The logic of our model is straightforward. In the rapid accretion regime, the funnel geometry produces strongly anisotropic emission: polar sightlines receive a brighter, harder continuum, while equatorial views are self-shadowed and depleted in XUV/soft-X-ray photons \citep{Wang2014,Lupi2024b,Madau2026}. 
Hard X-rays may be further weakened if the corona is efficiently Compton-cooled by the intense soft-photon field \citep{Madau2024,Trinca2026}. 
A dusty component of modest covering factor can then redden high-inclination sightlines, producing the rest-optical slopes and V-shaped continua of LRDs, while lower-inclination views appear as LBDs. Thus, the observed LRD/LBD differences arise mainly from orientation and line-of-sight processing, rather than from distinct engines or evolutionary phases \citep[e.g.,][]{Kido2025,Naidu2025,Pacucci2026}.

We adopt the clumpy line-of-sight interception formalism of Paper I; details are given in the Appendix. \textcolor{black}{As shown in Figure~\ref{fig:LRD_LF}, the observed LRD LF lies below the parent BLAGN UV LF. There is tentative evidence, at approximately $2.5\sigma$, that the LRD/BLAGN ratio in the brightest bin, $M_{1500}=-21$, is lower than at $M_{1500}=-19$; the decrease in the $M_{1500}=-20$ bin is not statistically significant.} In our framework, this offset reflects orientation-dependent dust interception: LRDs are the minority of \textcolor{black}{BLAGNs} viewed through an effective dusty obscurer, while less obscured sightlines appear as LBDs. \textcolor{black}{This possible bright-end decrease motivates exploring a luminosity-dependent dust covering factor}, as may be expected if radiation pressure reduces circumnuclear dust in more luminous systems \citep[e.g.,][]{Ricci2017,Ricci2022}.

Our implementation is a hybrid forward Monte Carlo model anchored to the empirical double-power-law UV LF of BLAGNs obtained by jointly fitting the spectroscopic samples of \citet{Taylor2025_BHMF} (\(3.5<z<6\)) and \citet{Juod2026} (\(4<z<7\)). We draw mock parent sources over the range \(-23\le M_{1500}\le -16.75\) mag using this best-fitting BLAGN LF, while the fit to the LRD LF itself is restricted to bins brighter than \(M_{1500}=-17\) mag. 

{\red We parameterize the parent LF as
\begin{equation}
\Phi_{\rm BLAGN}(M)=\frac{\Phi_\star}
{10^{0.4(\alpha+1)(M-M_\star)}+10^{0.4(\beta+1)(M-M_\star)}}.
\label{eq:parent_DPL}
\end{equation}
The weighted joint fit gives $\log_{10}(\Phi_\star/{\rm Mpc}^{-3}\,{\rm mag}^{-1})=-4.72\pm0.21$, $M_\star=-21.75\pm0.22$, and $\alpha=-1.63\pm0.16$. The bright-end slope reaches the adopted lower bound and is fixed at $\beta=-8$; its precise value has negligible effect over the luminosity range that dominates the forward model. The Taylor et al.\ point at $M_{1500}=-17$ is displayed but excluded from this double-power-law fit, since including its abrupt downturn would require an additional faint-end turnover. The shaded parent-LF intervals in Figure~\ref{fig:LRD_LF} propagate the covariance of the three fitted parameters.}

Each mock source is also assigned an accretion rate drawn from a Gaussian distribution in $\log\dot m$,\footnote{Here, we define $\dot m\equiv 0.1\dot M c^2/L_{\rm Edd}$.}, centered at $\langle\log\dot m\rangle=1.3$, corresponding to the midpoint of the super-Eddington model grid explored here. We keep this mean fixed and treat the dispersion $\sigma_{\log\dot m}$ as a free parameter, in analogy with the width of the dust-attenuation distribution (see below). This choice anchors the luminosity inversion to the representative accretion-rate scale of our models while allowing for source-to-source variation within the rapidly accreting parent population.


For each mock source we then assign an isotropically distributed viewing angle and evaluate the corresponding inclination-dependent anisotropic SED predicted by our tabulated super-Eddington models \citep{Madau2026}. The probability that the line of sight intersects circumnuclear dusty clouds is described by a luminosity-dependent covering factor,
\begin{equation}
C_{\rm dust}(M_{1500}) = \frac{C_0}{1 + \exp\left[-k(M_{1500}-M_{\rm break})\right]},
\label{eq:sigmoid}
\end{equation}
where $C_0$ is the faint-end plateau, $M_{\rm break}$ is the characteristic transition magnitude, and $k$ controls the sharpness of the decline toward the bright end. Since the present bright-end LRD counts only weakly constrain the transition width, we fix $k=4$ and treat $(C_0,M_{\rm break})$ as free parameters.

{\red It is important to distinguish this global covering factor from the inclination-dependent interception probability. At a given observed magnitude, $C_{\rm dust}(M_{1500})$ is the probability of dust interception averaged over an isotropic distribution of sightlines; it is not a second function evaluated at the inclination of an individual source. Equation~(\ref{eq:sigmoid}) fixes the equatorial normalization $N_{0,d}(M_{1500})$ by requiring that the solid-angle average in equation~(\ref{eq:Cdust}) equal $C_{\rm dust}(M_{1500})$. For a mock source whose inclination $i$ has already been drawn, the probability of intercepting at least one cloud is then
\begin{equation}
\begin{aligned}
1-P_{{\rm esc},d}(i\mid M_{1500})
&=1-\exp\!\left\{-N_{0,d}(M_{1500})\right.\\
&\hspace{2.7cm}\left.\times
\exp\!\left[-\frac{\cos^2 i}{2\sigma_d^2}\right]\right\}.
\end{aligned}
\label{eq:Pint_M_i}
\end{equation}
Thus, luminosity sets the population-averaged covering factor, while inclination determines how that probability is distributed over sightlines. The two dependencies are complementary rather than degenerate definitions of the same quantity: at fixed $M_{1500}$, equatorial views have the largest interception probability, whereas the normalization averaged over all $i$ is exactly $C_{\rm dust}(M_{1500})$.}

Given $C_{\rm dust}$, we determine probabilistically whether the source intercepts the dusty obscurer using the same clumpy Poisson formalism introduced in Paper I. The dusty cloud population is assumed to have angular width $\sigma_d=0.4$, corresponding to a geometrically thick equatorial distribution whose normalization is fixed by equation~(\ref{eq:sigmoid}). {\red We adopt a gray AGN-type extinction curve, qualitatively similar to the flat attenuation laws inferred for reddened active nuclei \citep[e.g.,][]{Gaskell2004}. We fix $R_V=5.5$ and suppress the 2175-\AA\ feature ($B=0$), giving $A_{1500}/A_V\simeq1.30$. The extinction per obscuring cloud is drawn from a truncated normal distribution at $A_V^c>0$ with fixed mean $\langle A_V^c\rangle=3.0$\,mag and free dispersion $\sigma_{A_V^c}$. We also include a uniform host-galaxy screen of $A_V^{\rm host}=0.5$\,mag applied to all sightlines. Thus, an unobscured LBD-like sightline receives only the baseline host attenuation, whereas a one-cloud dust-intercepted sightline has a characteristic total extinction of $A_V\simeq3.5$\,mag. The total transmission along dust-intercepted sightlines is computed from the conditional Poisson cloud-crossing formalism described in the Appendix; because more than one cloud may be intersected, their effective attenuation is not represented by a single fixed total $A_V$.

At fixed values of the remaining model parameters, a sensitivity scan favors $R_V\simeq5.5$; steeper curves with $R_V=4.6$--5.1 tend to overproduce color-selected LRDs. Re-optimizing the LF parameters partially compensates for this effect through a smaller fitted covering factor, while still grayer curves with $R_V\gtrsim6$ provide statistically comparable fits only by requiring substantially larger covering factors. The LRD LF therefore does not constrain $R_V$ independently of the dust geometry. We retain $R_V=5.5$ as a representative gray AGN attenuation curve and treat the dependence on $R_V$ as a systematic uncertainty rather than introducing it as an additional free parameter.}

\begin{figure}[!hbt]
\centering
\includegraphics[width=\hsize,trim=0 0 0 0,clip]{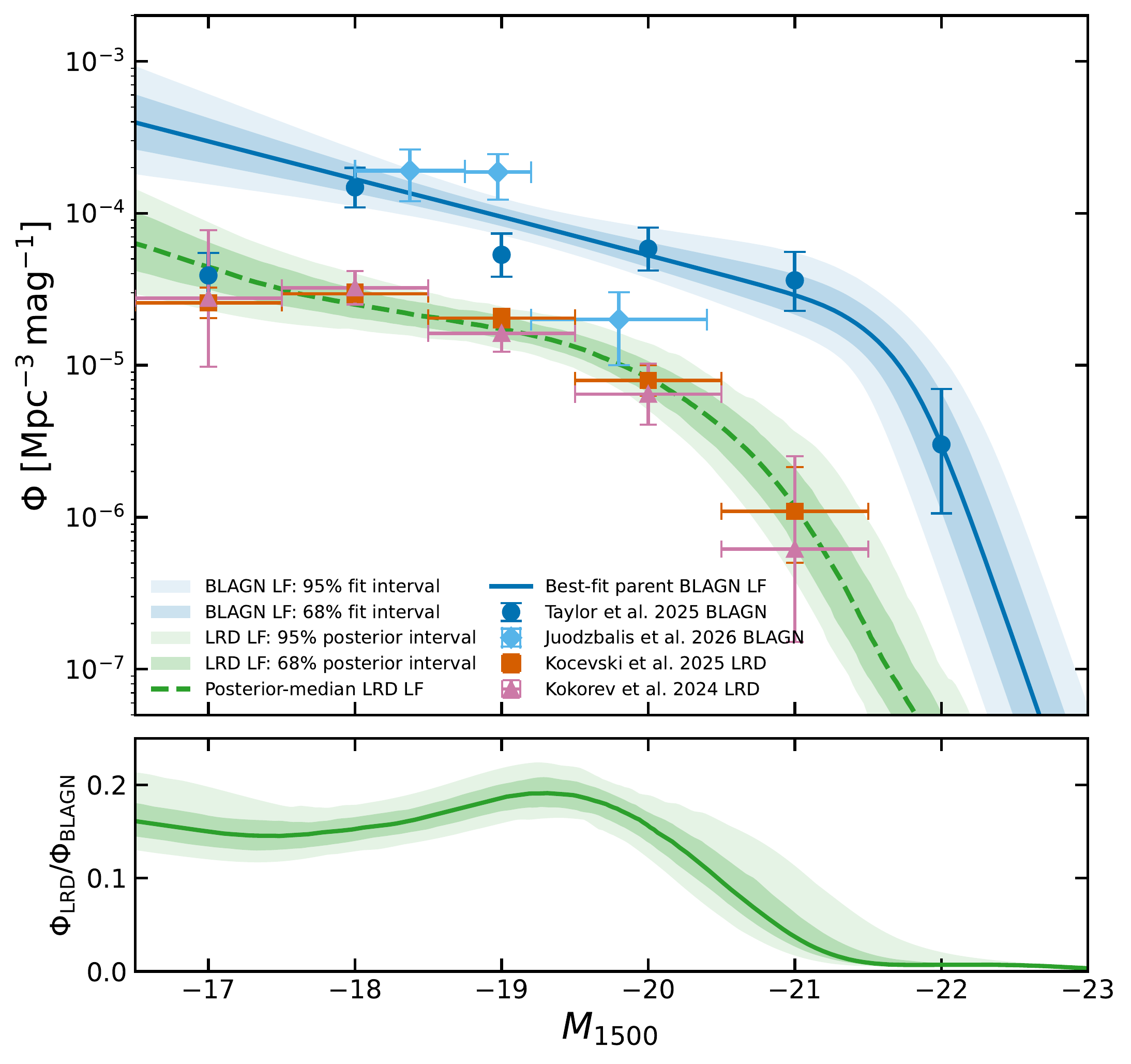}
\caption{Observed and model UV LFs of LRDs and their parent BLAGN population. The upper panel shows the double-power-law parent BLAGN UV LF fitted to the spectroscopic measurements of \citet{Taylor2025_BHMF} and \citet{Juod2026}, and the photometric LRD LFs of \citet{Kocevski2025} and \citet{Kokorev2024}. The green dashed curve is the posterior-median LRD LF. Dark and light shaded regions show the 68\% and 95\% intervals, respectively; the BLAGN bands propagate the covariance of the parent-LF fit, while the LRD bands additionally propagate the posterior uncertainty of the forward model. The lower panel shows the corresponding apparent ratio $\Phi_{\rm LRD}/\Phi_{\rm BLAGN}$. {\red The best-fitting solution has $\chi^2/{\rm dof}=1.01$ for 10 data points and four free parameters, with $\sigma_{\log\dot m}=0.06$, $\sigma_{A_V^c}=0.05$, $C_0=0.26$, and $M_{\rm break}=-20.17$. The remaining parameters are fixed to $\langle\log\dot m\rangle=1.3$, $\langle A_V^c\rangle=3.0\,{\rm mag}$, $\sigma_d=0.4$, $A_V^{\rm host}=0.5\,{\rm mag}$, $R_V=5.5$, $B=0$, and $k=4$. The implied number-density-weighted mean covering factor is $\langle C_{\rm dust}\rangle=0.23$, and $15\%$ of the modeled BLAGNs satisfy the adopted photometric LRD selection.}}
\label{fig:LRD_LF}
\end{figure}

At this stage, each mock source has a specified target observed magnitude $M_{1500}$, accretion rate $\dot m$, viewing angle, and total line-of-sight attenuation. We then determine the black hole mass iteratively by requiring that the final attenuated SED reproduce exactly the assigned observed UV magnitude at 1500\,\AA. In this way, the empirical BLAGN UV LF is preserved by construction, while dust-reddened sightlines are naturally mapped onto intrinsically more luminous and typically more massive systems than unobscured ones at the same observed $M_{1500}$.

From the final attenuated continuum we measure the rest-frame UV slope $\beta_{\rm UV}$ using windows centered at 1500, 2000, and 2500\,\AA, and the optical slope $\beta_{\rm opt}$ using windows at 4500, 6000, and 8000\,\AA. A mock source is classified as a photometric LRD if
\begin{equation}
\beta_{\rm opt}>0,
\qquad
-2.8<\beta_{\rm UV}< -0.37,
\end{equation}
following \citet{Kocevski2025}. The predicted LRD LF is then obtained from the fraction of mock BLAGNs in each magnitude bin that satisfy these criteria, multiplied by the fixed parent BLAGN LF.

\begin{figure*}[!hbt]
\centering
\includegraphics[width=0.95\hsize,trim=0 0 0 0,clip]{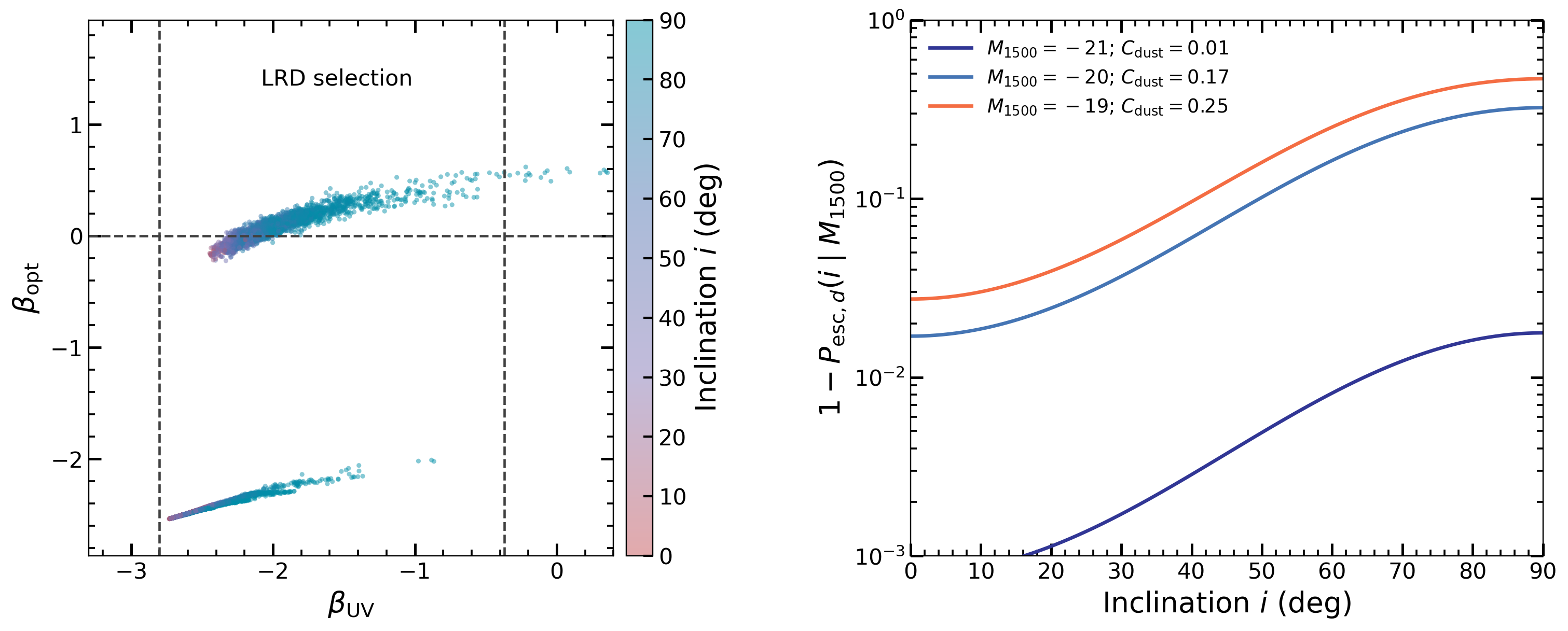}
\caption{\textcolor{black}{Orientation dependence of the best-fitting forward model. Left: synthetic rest-frame continuum slopes for the mock BLAGN population, color-coded by inclination. Dashed lines delimit the adopted photometric LRD selection. Right: probability that a sightline intercepts at least one dusty cloud, $1-P_{{\rm esc},d}(i\mid M_{1500})$, as a function of inclination at three observed UV magnitudes. A logarithmic ordinate is used to display the low interception probabilities of the brightest sources. The value of $C_{\rm dust}$ in each legend entry is the solid-angle average of the corresponding curve, weighted by $\sin i$. The luminosity-dependent global normalization and the inclination-dependent interception probability are therefore distinct but connected quantities.}}
\label{fig:orientation}
\end{figure*}

{\red Figure~\ref{fig:orientation} makes the orientation dependence explicit. The left panel shows the synthetic continuum slopes of the mock catalog, color-coded by inclination. The LRD selection preferentially isolates high-inclination objects: for the best-fitting realization the median inclination is $59.8^\circ$ for the full isotropic mock, $73.6^\circ$ for dust-intercepted sightlines, and $78.9^\circ$ for photometrically selected LRDs. The right panel shows that $C_{\rm dust}(M_{1500})$ controls the solid-angle normalization, while the interception probability at every luminosity rises strongly toward edge-on sightlines.}

\begin{figure*}[!hbt]
\centering
\includegraphics[width=0.95\hsize,trim=0 0 0 0,clip]{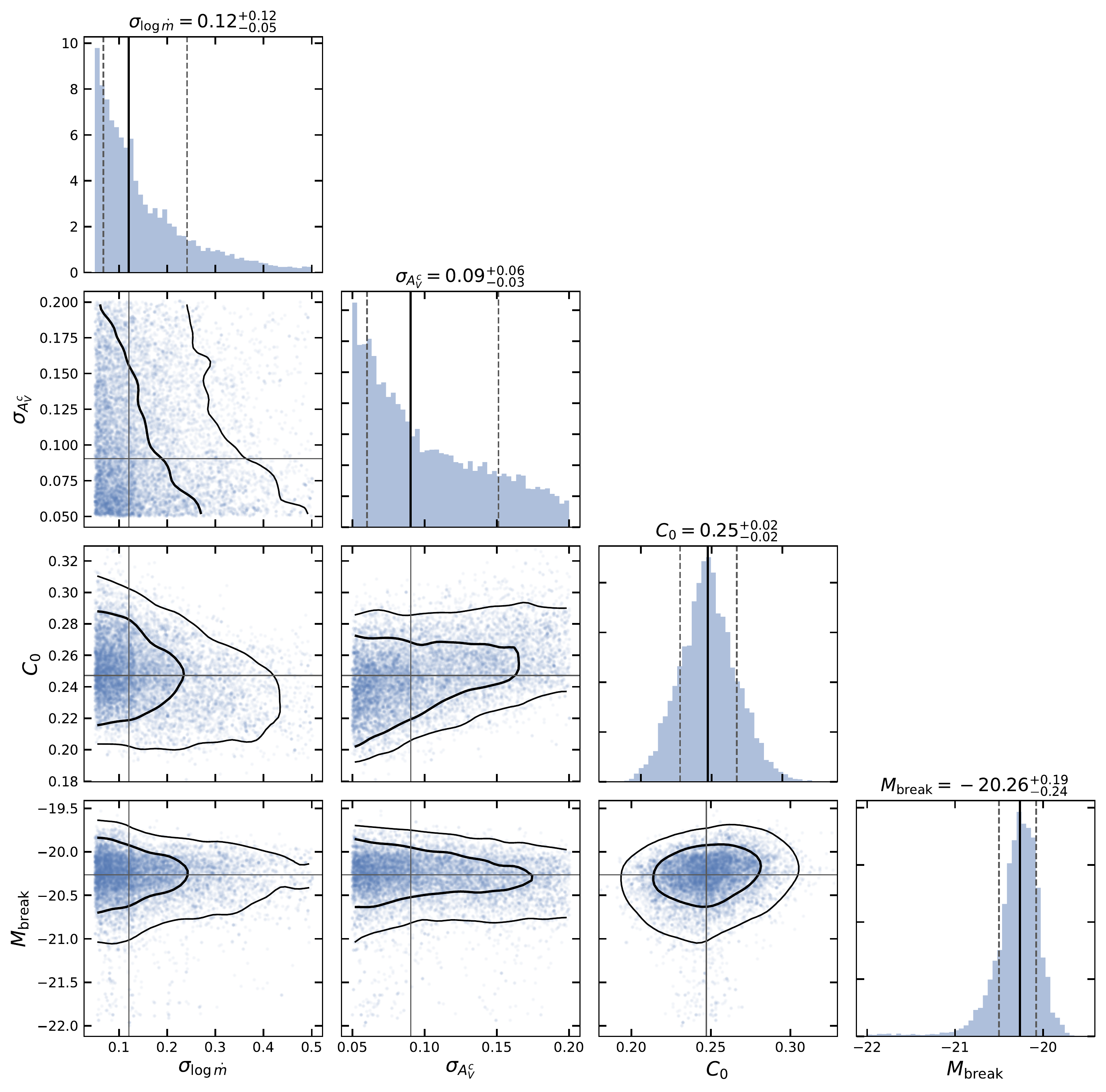}
\caption{\textcolor{black}{Posterior distributions for the four free parameters. Lines mark the medians and 16th--84th percentile intervals; contours enclose 68\% and 95\% probability. The best fit has $\chi^2/{\rm dof}=6.08/6=1.01$. The two dispersion parameters remain weakly constrained toward their lower prior limits.}}
\label{fig:corner}
\end{figure*}

The free parameters of the fiducial model are therefore $\sigma_{\log\dot m}$, $\sigma_{A_V^c}$, and the two sigmoid parameters $(C_0,M_{\rm break})$. The mean accretion rate, $\langle\log\dot m\rangle=1.3$, the mean per-cloud extinction,
{\red $\langle A_V^c\rangle=3.0\,{\rm mag}$, the dust angular width, $\sigma_d=0.4$, the host reddening screen, $A_V^{\rm host}=0.5\,{\rm mag}$, the attenuation-curve parameters, $R_V=5.5$ and $B=0$,} and the sigmoid sharpness, $k=4$, are held fixed.
{\red We sample the four-parameter posterior using bounded uniform priors over the ranges listed in Table~\ref{tab:model_parameters}. The resulting marginalized constraints are also listed in the table and shown in Figure~\ref{fig:corner}; the two dispersion parameters remain weakly constrained toward their lower prior bounds. All final model predictions and confidence bands are evaluated using 40,000 mock sources.}

\begin{table}[!hbt]
\centering
\color{black}
\caption{Model parameters. Uncertainties are the 16th--84th posterior percentiles.}
\label{tab:model_parameters}
\begin{tabular}{lccc}
\hline\hline
Parameter & Status & Prior/range & Result \\
\hline
$\sigma_{\log\dot m}$ & Free & $[0.05,0.5]$ & $0.12_{-0.05}^{+0.12}$ \\
$\sigma_{A_V^c}$ & Free & $[0.05,0.2]$ & $0.09_{-0.03}^{+0.06}$ \\
$C_0$ & Free & $[0.05,0.95]$ & $0.25_{-0.02}^{+0.02}$ \\
$M_{\rm break}$ & Free & $[-22,-16]$ & $-20.26_{-0.24}^{+0.19}$ \\
$\langle\log\dot m\rangle$ & Fixed & --- & $1.30$ \\
$A_V^{\rm host}$ & Fixed & --- & $0.5$ mag \\
$\langle A_V^c\rangle$ & Fixed & --- & $3.0$ mag \\
$R_V$ & Fixed & --- & $5.5$ \\
$B$ & Fixed & --- & $0$ \\
$\sigma_d$ & Fixed & --- & $0.40$ \\
$k$ & Fixed & --- & $4.0$ \\
\hline
\end{tabular}
\end{table}

\begin{figure}[!hbt]
\centering
\includegraphics[width=\hsize,trim=0 0 0 0,clip]{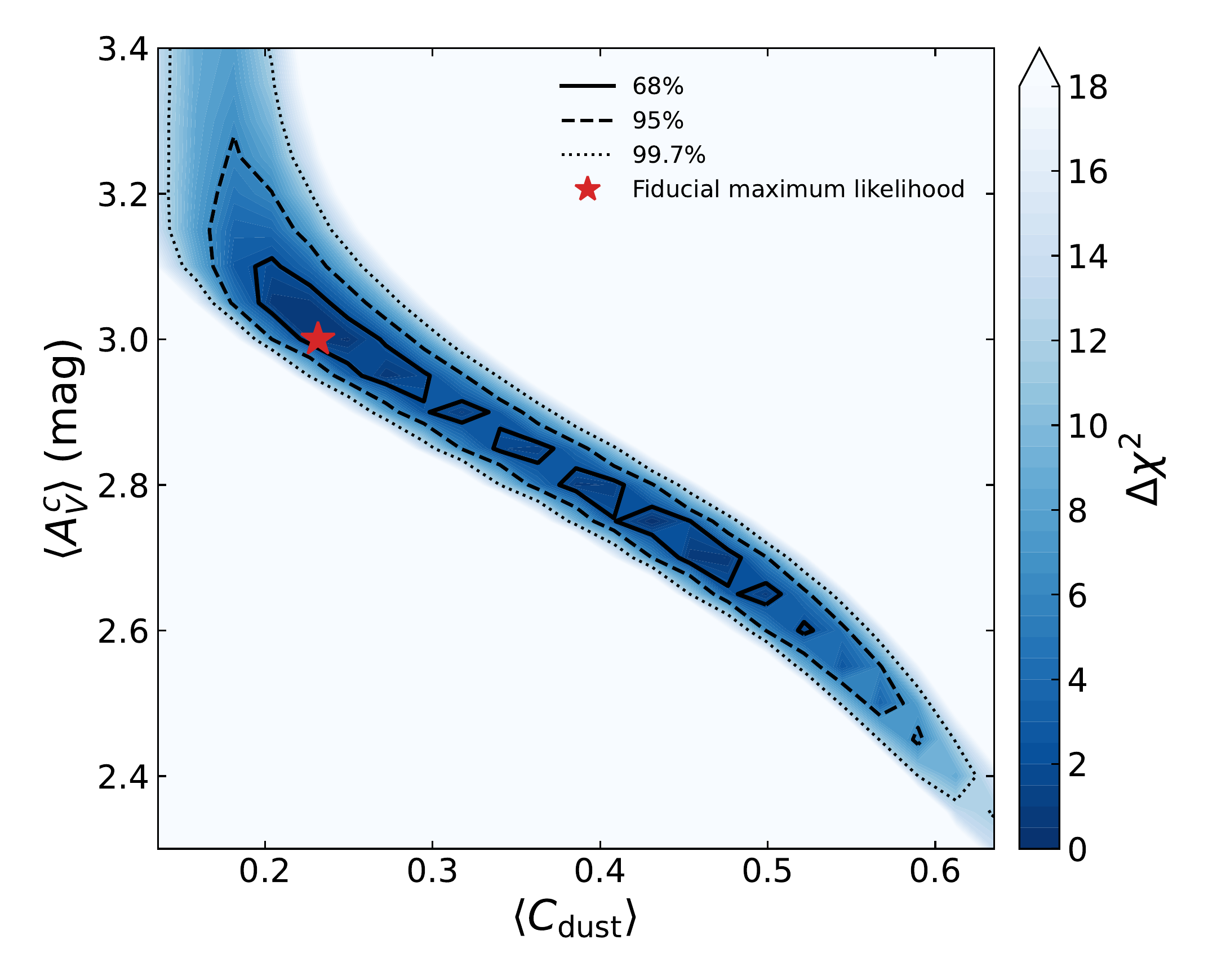}
\caption{Conditional profile likelihood in the $(\langle C_{\rm dust}\rangle,\langle A_V^c\rangle)$ plane. At each $(\langle A_V^c\rangle,C_0)$ grid point, we minimize over $\sigma_{A_V^c}$; $\sigma_{\log\dot m}=0.063$ and $M_{\rm break}=-20.17$ are held fixed at their fiducial maximum-likelihood values. Each value of $C_0$ is converted to the corresponding number-density-weighted mean covering factor. Solid, dashed, and dotted contours denote the 68\%, 95\%, and 99.7\% confidence regions. The red star marks the fiducial four-parameter maximum-likelihood solution, for which $C_0=0.255$, $\langle C_{\rm dust}\rangle=0.232$, and $\langle A_V^c\rangle=3.0$~mag.}
\label{fig:dust_degeneracy}
\end{figure}

The posterior constraints are conditional on the fixed characteristic cloud extinction and attenuation curve. To quantify this systematic dependence, we performed a conditional profile-likelihood scan in the $(\langle A_V^c\rangle,C_0)$ plane, minimizing over $\sigma_{A_V^c}$ at each grid point while holding $\sigma_{\log\dot m}$ and $M_{\rm break}$ at their maximum-likelihood values. Figure~\ref{fig:dust_degeneracy} shows a pronounced degeneracy: a lower characteristic extinction per cloud can be compensated by a larger dust covering factor, which both increases the fraction of intercepted sightlines and, in the adopted Poisson clumpy model, increases the mean number of clouds conditional on interception. The fiducial four-parameter maximum-likelihood solution lies within the 68\% confidence region. Thus, the LRD LF does not determine the extinction and covering factor independently, and the narrow posterior interval on $C_0$ should be interpreted as conditional on the adopted dust prescription.

\section{Results and discussion}
\label{sec:results}

Before presenting the LF comparison, we briefly summarize the logic of the population synthesis:

\begin{enumerate}

\item We adopt the observed rest-frame UV LF of \textcolor{black}{BLAGNs} at $z\gtrsim4$ as the parent distribution of the little-dot population. In this framework, the Taylor et al.\ and Juod\v{z}balis et al.\ BLAGN LF provides the underlying UV number density from which both unobscured LBDs and dust-reddened LRDs are generated through orientation and circumnuclear obscuration.

\item Each Monte Carlo source is assigned a target observed UV magnitude
$M_{1500}$ drawn from this empirical BLAGN LF, 
together with an accretion rate $\dot m$ drawn from a lognormal distribution with fixed center $\langle\log\dot m\rangle=1.3$ and free dispersion $\sigma_{\log\dot m}$.

\item We then assign each source a random isotropic viewing angle and evaluate the corresponding inclination-dependent anisotropic SED predicted by the accretion-flow model. The viewing angle changes both the observed continuum normalization and the UV-to-optical spectral shape, as expected for a geometrically thick super-Eddington emitter.

\item We probabilistically determine whether the line of sight intercepts circumnuclear dusty clouds. The interception probability is described by a luminosity-dependent covering factor $C_{\rm dust}(M_{1500})$, modeled as a sigmoid that decreases toward the bright end. This luminosity dependence is required because LRDs are rare among the most UV-luminous \textcolor{black}{BLAGNs}; a luminosity-independent covering factor would overproduce bright LRDs.

\item For dust-intercepted sightlines, the inclination-dependent SED is attenuated with a gray extinction law. {\red The per-cloud extinction is drawn from a distribution centered on $\langle A_V^c\rangle=3.0$~mag, and a foreground host-galaxy screen with $A_V^{\rm host}=0.5$~mag is applied to all sightlines.} Dust-free sightlines therefore retain only the foreground attenuation, whereas obscured sightlines acquire substantially larger UV suppression.

\item Given the total line-of-sight attenuation and the chosen inclination, we determine the black-hole mass iteratively by requiring that the final attenuated SED reproduce exactly the target observed UV magnitude $M_{1500}$. In this way, the mock catalog preserves by construction the observed BLAGN UV LF while allowing dust-reddened sightlines to correspond to intrinsically more luminous and typically more massive systems than unobscured ones at the same observed $M_{1500}$.

\item From each final mock SED we measure the synthetic UV and optical continuum slopes and classify as LRDs only those objects satisfying the adopted photometric color-selection criteria. Counting these selected mock LRDs in each magnitude bin yields the predicted UV LRD LF.

\item The same mock SEDs are also used to compute rest-frame optical magnitudes at 4500 \AA\ and 6500 \AA. Because dust-reddened, high-inclination sources have much redder optical-to-UV colors than unobscured BLAGNs, objects that are only moderately bright in observed $M_{1500}$ can populate substantially brighter optical-magnitude bins. This color migration increases the apparent LRD/BLAGN ratio in the optical relative to the UV.

\end{enumerate}

\subsection{A low covering solution}

Figure~\ref{fig:LRD_LF} compares the resulting forward-model predictions with the observed UV LF measurements. The parent BLAGN LF is described by a fixed double power law (blue solid curve), while the predicted LRD LF (green dashed curve) is obtained by applying orientation-dependent dust interception and the LRD color selection to this parent population. The model reproduces the observed normalization and overall shape of the LRD LF over $-21\lesssim M_{1500}\lesssim -17$. As shown in the lower panel, it also predicts a strongly luminosity-dependent apparent LRD fraction, $\Phi_{\rm LRD}/\Phi_{\rm BLAGN}$, rising from below
$1\%$ at $M_{1500}\sim -22$ to a maximum of $\sim20\%$ near $M_{1500}\sim -19$, with the subsequent faint-end behavior depending on the
extrapolated BLAGN LF and the magnitude range sampled in the Monte Carlo draw. This behavior reflects the combined effects of a dust covering factor that declines toward high luminosities and a color-selection efficiency that depends on luminosity, inclination, and attenuation. The same mock catalog predicts a substantially larger apparent LRD/BLAGN ratio when the population is selected in the rest-frame optical rather than in the UV; this additional wavelength dependence is discussed below.

{\red The fiducial solution corresponds to a number-density-weighted mean dust covering factor of $\langle C_{\rm dust}\rangle \simeq 0.23$, implying that about one quarter of BLAGN sightlines are dust-intercepted over the modeled luminosity range. In the mock population, $\simeq15\%$ of BLAGNs satisfy the photometric LRD criterion, while an additional $\simeq8\%$ are dust-intercepted but fall outside the adopted LRD selection box. Thus about two-thirds of the obscured sightlines are classified as LRDs, indicating that the adopted gray dust prescription is effective at producing LRD-like continua for a substantial fraction of dust-intercepted sources. The maximum-likelihood solution favors narrow distributions, with $\sigma_{\log\dot m}=0.06$ and $\sigma_{A_V^c}=0.05$; after marginalizing over the covering-factor parameters, the posterior medians are $0.12$ and $0.09$, respectively, with both dispersions remaining weakly constrained toward their lower prior bounds.}

\subsection{Predicted wavelength dependence of the apparent LRD fraction}

Table~\ref{tab:opt_UV_ratios} compares the best-fitting apparent ratio $\Phi_{\rm LRD}/\Phi_{\rm BLAGN}$ when the \textcolor{black}{modeled BLAGN population} is binned by observed rest-frame magnitude at $1500$, $4500$, and 6500 \AA. The UV-selected ratio rises from $3\%$ at $M_{1500}=-21$ to $20\%$ at $M_{1500}=-19$, consistent with the obscured minority fraction inferred from the fiducial dust covering model. In contrast, the corresponding optical-selected ratios are systematically larger: the 4500 \AA\ ratio reaches $27\%$ at $M_{4500}=-20$, while the 6500 \AA\ ratio is larger still, $37\%$ at 
$M_{6500}=-20$.

\begin{table}
\centering
\caption{Predicted apparent LRD/BLAGN ratios at representative UV and optical magnitudes.}
\label{tab:opt_UV_ratios}
\begin{tabular}{cccc}
\hline
Magnitude & 1500 \AA & 4500 \AA & 6500 \AA \\
\hline
$-21$ & 0.03 & 0.17 & 0.39 \\
$-20$ & 0.15 & 0.27 & 0.37 \\
$-19$ & 0.20 & 0.29 & 0.33 \\
\hline
\end{tabular}
\end{table}

This behavior does not imply that a larger fraction of lines of sight intrinsically produce LRDs at longer wavelength. In the model, the true fraction of dust-intercepted sightlines remains fixed by the adopted circumnuclear cloud covering factor and is of order $\sim20\%$. The larger optical ratios arise because the quantity plotted is not the intrinsic sightline fraction, but the ratio of {observed LFs} evaluated at fixed observed magnitude in each band.
For each mock source, the \textcolor{black}{parent BLAGN population} is sampled from the observed UV LF, and the emergent SED is then computed for its assigned inclination and dust interception. Dust attenuation shifts the same object by an amount
\[
M_{\lambda,{\rm obs}} = M_{\lambda,{\rm int}} + A_\lambda,
\]
with a wavelength dependence that is highly non-uniform. At fixed observed $M_{1500}$, a dust-intercepted source must be intrinsically more luminous than an unobscured source in order to compensate for the UV attenuation.
Although the adopted extinction curve is gray compared with a standard Milky-Way law, the attenuation still decreases from 1500 to 4500 and 6500\,\AA. The same reddened objects therefore appear relatively brighter in the rest-frame optical than in the UV, shifting them into brighter optical-magnitude bins and increasing the apparent LRD/BLAGN ratio at optical wavelengths.

The consequence is a color-driven migration across magnitude bins. An intrinsically luminous dusty source that appears extremely faint in the UV can still populate a moderate optical luminosity bin, whereas the unobscured BLAGN population remains concentrated closer to its intrinsic UV luminosity ranking. Rest-frame optical bins are therefore populated efficiently by red, high-inclination objects that would be strongly underrepresented in a UV-selected comparison. The apparent ratio $\Phi_{\rm LRD}/\Phi_{\rm BLAGN}$ at fixed optical magnitude can thus substantially exceed the underlying geometric obscured fraction, even though the number of LRD-producing lines of sight is unchanged.

This effect becomes progressively stronger toward longer wavelength because $A_\lambda$ decreases monotonically from the UV to the optical. The same dust-reddened subset that is heavily suppressed at 1500\,\AA\ is only moderately shifted at 4500 and less shifted still at 6500\,\AA, causing the observed LRD counts to be increasingly recovered relative to the blue BLAGN denominator. The systematic rise of the ratio from UV to optical wavelengths is therefore a direct consequence of wavelength-dependent extinction combined with magnitude binning in the observed frame.

The larger ratios predicted at  4500 \AA\ and 6500 \AA\ provide a direct
observational test of the orientation-based unification scenario. If LRDs are indeed dust-intercepted members of the same \textcolor{black}{BLAGN population}, rest-frame optical surveys should recover a substantially larger apparent LRD fraction than UV-selected samples over comparable observed luminosity ranges.

\begin{figure}[!ht]
\centering
\includegraphics[width=0.95\hsize,trim=0 0 0 0,clip]{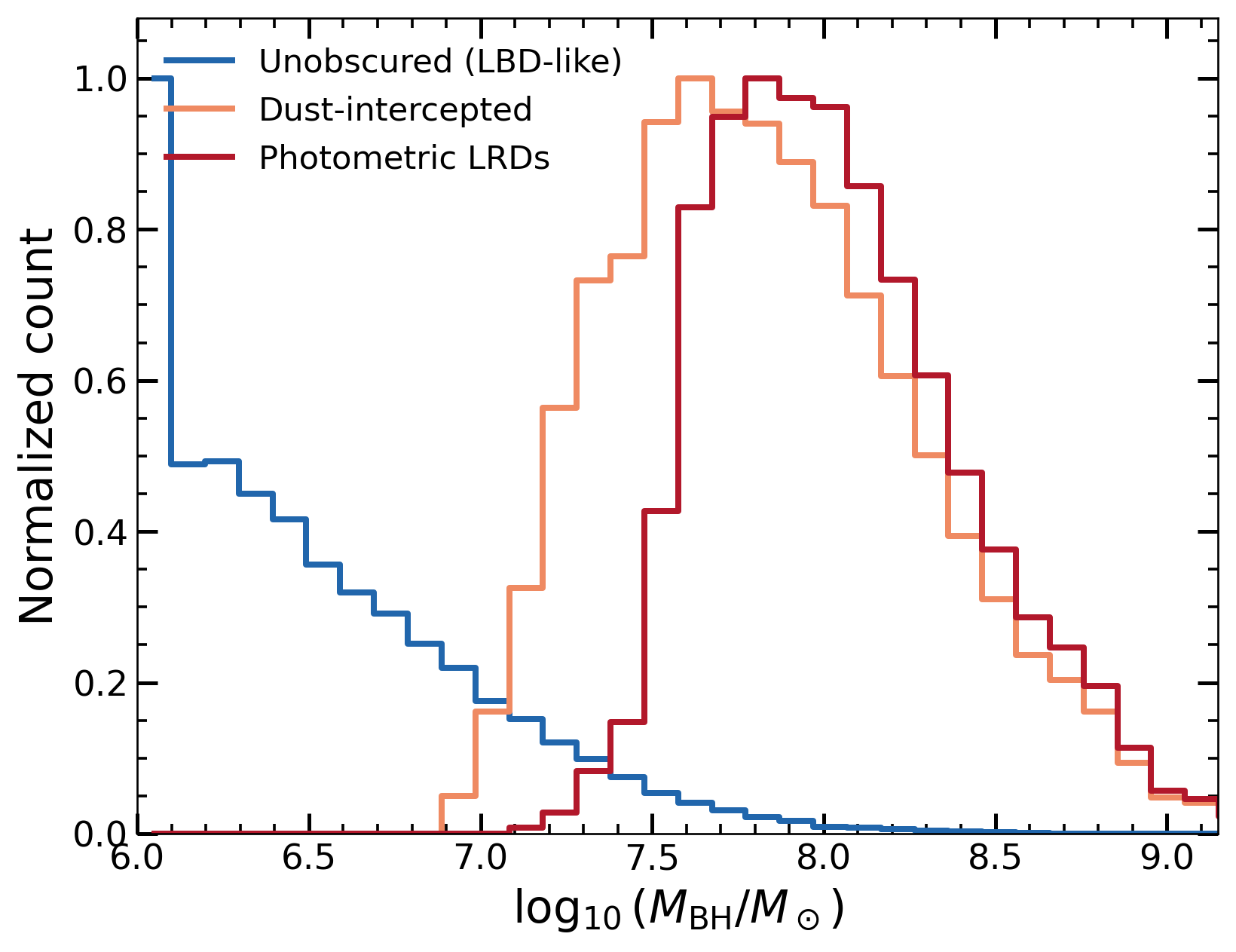}
\includegraphics[width=0.95\hsize,trim=0 0 0 0,clip]{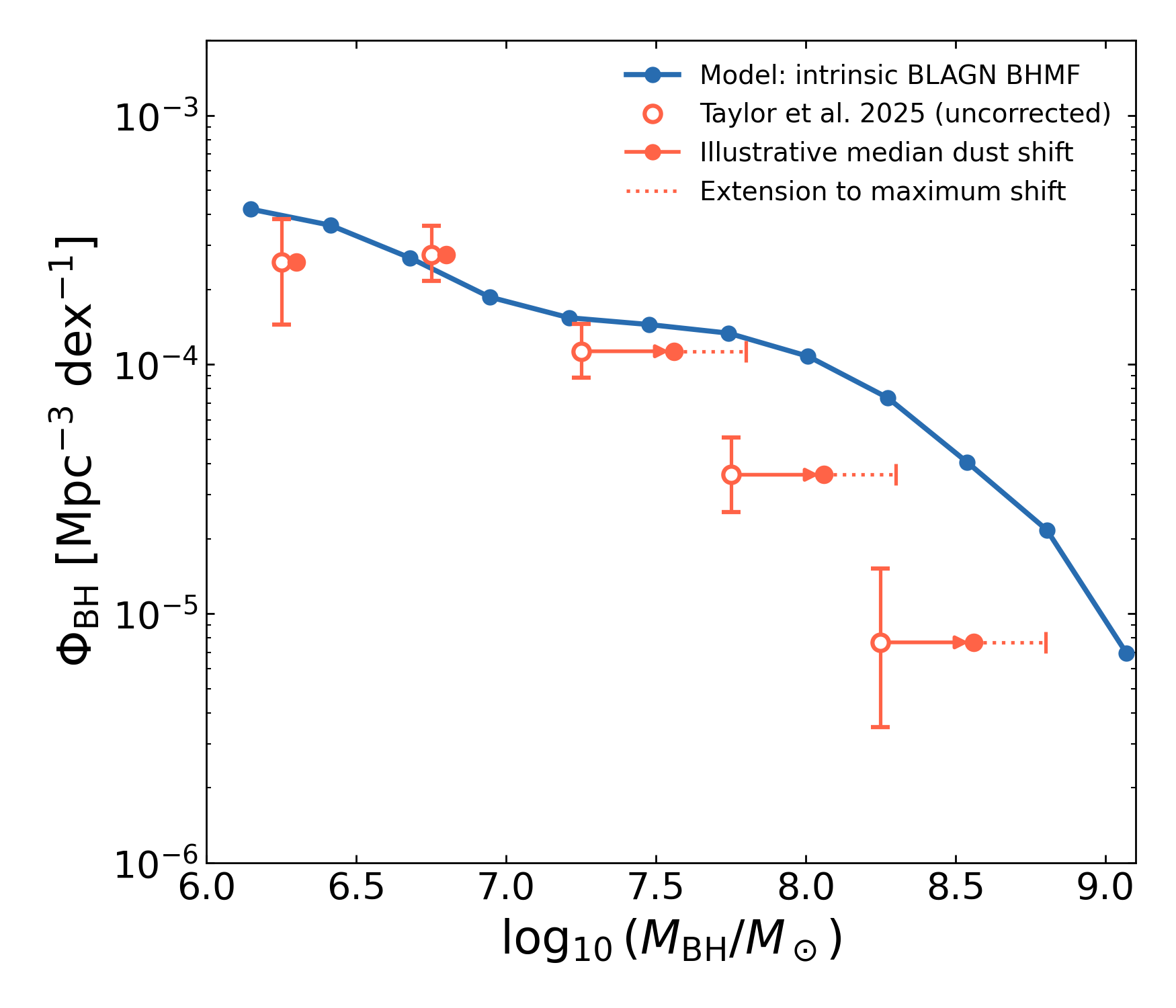}
\caption{Black-hole mass demographics implied by the best-fitting forward model. {\it Top panel:} normalized inferred mass distributions of unobscured (LBD-like) sightlines, all dust-intercepted sources, and the subset classified as photometric LRDs. {\it Bottom panel:} intrinsic BLAGN black-hole mass function implied by the same mock population after the iterative inversion from the observed BLAGN UV LF. Open red points show the BHMF inferred by \citet{Taylor2025_BHMF} from uncorrected broad-H$\alpha$ virial masses. Filled red points illustrate the horizontal shifts obtained by applying their reported median mass corrections: $+0.05$ dex below $10^7\,M_\odot$ and $+0.31$ dex above it. For the massive bins, dotted extensions reach the largest correction in their sample, $+0.55$ dex. These shifts leave the plotted number densities unchanged and are not a recomputed extinction-corrected BHMF.}
\label{fig:BHMF}
\end{figure}

\subsection{Black-hole mass demographics}

Figure~\ref{fig:BHMF} summarizes the black-hole mass demographics implied by the best-fitting forward model. The top panel compares the normalized inferred mass distributions of unobscured sightlines (LBD analogs), all dust-intercepted sources, and the subset of dust-intercepted sources that additionally satisfy the photometric LRD color criteria.

The inferred black-hole mass distributions of unobscured and dust-reddened mock sources differ markedly. Non-dusty sightlines are concentrated toward the low-mass end of the allowed SED grid, since only modest intrinsic UV luminosities are required to reproduce the assigned observed $M_{1500}$ in the absence of strong attenuation. By contrast, dust-intercepted sightlines are shifted to substantially larger inferred masses, peaking near $\log (M_{\rm BH}/M_\odot)\sim 7.5$ and extending to beyond $10^{8.5}\,M_\odot$, while the subset that additionally satisfies the photometric LRD color criteria is displaced even further toward the high-mass tail.

This pronounced mass segregation should not be interpreted as evidence that LRDs and LBDs arise from intrinsically distinct parent populations. In the model, both classes are drawn from the same \textcolor{black}{BLAGN UV LF} and differ only by orientation and line-of-sight obscuration. The separation in inferred $M_{\rm BH}$ is instead a direct selection effect produced by the iterative luminosity inversion. Once the target observed $M_{1500}$ is held fixed, a dust-reddened source must correspond to a substantially brighter intrinsic UV continuum than an unobscured source in order to remain detectable within the same observed UV magnitude bin. This forces the obscured solutions toward larger black-hole masses. Conversely, dust-intercepted BLAGNs with intrinsically lower masses would be attenuated below the UV luminosity range sampled by the present data and would therefore not enter the observed LRD population.

The lower panel shows the black-hole mass function (BHMF) implied by the same
best-fitting mock \textcolor{black}{BLAGN parent population}. Over the range
$10^{6}\lesssim M_{\rm BH}/M_\odot\lesssim10^{9}$, the inferred BHMF declines monotonically, with a progressively steeper fall toward the highest masses. This behavior reflects the bright-end cutoff of the parent \textcolor{black}{BLAGN UV LF}, together with the adopted mapping between UV luminosity, accretion rate, and black-hole mass. It should therefore not be interpreted as a direct measurement of the intrinsic cosmic AGN BHMF, but rather as the mass distribution of actively accreting \textcolor{black}{BLAGNs selected through the observed UV LF} under the best-fitting accretion-rate distribution.

{\red For comparison, Figure~\ref{fig:BHMF} also shows the BLAGN BHMF inferred by
\citet{Taylor2025_BHMF} using uncorrected broad-H$\alpha$ virial masses. They
estimated median extinction corrections of $+0.05$ dex below $10^7\,M_\odot$
and $+0.31$ dex above it, with a maximum correction of $+0.55$ dex for their
most reddened source. We illustrate these corrections as horizontal shifts at
fixed number density; because the individual mass posteriors and completeness
weights are unavailable, the shifted points should not be interpreted as a
recomputed extinction-corrected BHMF. The comparison therefore shows only the
direction and approximate magnitude of the mass correction: extinction shifts
the Taylor et al.\ estimates toward higher masses, while the corresponding
change in number density cannot be inferred from the published binned
measurements.}

{\red To characterize the high-mass tail directly, we examined the mock sources
with $\log(M_{\rm BH}/M_\odot)\geq8$. Of these, 94\% are viewed through the
dusty cloud distribution and 90\% satisfy the adopted LRD selection. Their
median observed and intrinsic UV magnitudes are $M_{1500}=-18.8$ and $-23.5$,
respectively; for the LRD subset the corresponding values are $-18.7$ and
$-23.6$. The high-mass excess is therefore associated primarily with objects
that are present at moderate observed UV luminosities but are shifted to much
brighter intrinsic luminosities, and hence to larger inferred masses, by the
dust correction.}

{\red Because the inversion from UV luminosity to black-hole mass remains
partially degenerate with the adopted form of $p(\log\dot m)$, we repeated the
fit at fixed $\langle\log\dot m\rangle=0.67$, 0.85, 1.09, 1.30, and 1.50,
holding $\sigma_{\log\dot m}=0.06$ fixed and reoptimizing
$\langle A_V^c\rangle$, $\sigma_{A_V^c}$, $C_0$, and $M_{\rm break}$. The
refitted mean cloud extinction remains in the narrow range
$\langle A_V^c\rangle=2.8$--3.0 mag, while the number-density-weighted mean
covering factor spans only $\langle C_{\rm dust}\rangle=0.25$--0.29. The solutions through
$\langle\log\dot m\rangle=1.30$ retain acceptable LRD-LF fits, with
$\chi^2/{\rm dof}=0.9$--1.3, while the 1.50 case is less satisfactory
($\chi^2/{\rm dof}=1.9$). As expected, decreasing the mean accretion rate
shifts the inferred mass distribution and BHMF toward larger masses, whereas
the 1.50 model shifts them downward. None of these conditional fits, however,
removes the high-mass offset relative to the Taylor et al.\ determination.
The high-mass tail remains sensitive to the assumed accretion-rate
distribution, the dust correction, and the limited dynamic range of the
present BLAGN LF, and should therefore be regarded as indicative rather than
definitive. We finally reiterate that the inferred BHMF is for actively
accreting BLAGNs that are individually detectable in current spectroscopic
samples. Fainter accreting black holes, such as those inferred from stacked
analyses \citep[e.g.,][]{Geris2026}, would preferentially add low-mass systems
and therefore steepen the faint-/low-mass end of the BHMF.}

{\red The preceding results assume that the rest-frame UV continuum is
AGN-dominated. Under this assumption, the same nuclear component must reproduce
both the blue UV slope and the red optical--near-IR continuum of LRDs. Strong
obscuration by a standard extinction curve would suppress and redden the AGN
far-UV emission too strongly; an AGN-dominated interpretation therefore
requires a comparatively gray attenuation law, as adopted in our fiducial
model. This assumption also ties the observed $M_{1500}$ directly to the
attenuated AGN luminosity and hence underlies the luminosity inversion used to
infer the black-hole mass.}

{\red If the observed UV continuum is instead dominated by an unobscured
stellar component, the blue UV slope no longer constrains the extinction law
toward the nucleus, and the AGN can be obscured by a more conventional,
non-gray dust screen. A nuclear extinction of a few magnitudes can then
strongly suppress the AGN in the far UV, leaving the unobscured stellar
component to produce the observed blue UV spectrum, while the obscured AGN
re-emerges at optical and near-IR wavelengths and powers the broad H$\alpha$
emission. Our modeling of LRDs shows that, after adding an unobscured stellar
component that dominates at 1500\,\AA, the optical--near-IR SED still requires
$A_V^c\simeq3$ mag toward the AGN (Madau \& Maiolino, in preparation).}

{\red In this stellar-dominated limit, the observed $M_{1500}$ primarily traces
the host rather than the attenuated nuclear luminosity. Nuclear obscuration
therefore produces less migration between UV-magnitude bins, and the mapping
from the parent BLAGN LF to the LRD LF depends on the joint stellar--AGN
luminosity distribution, the AGN-to-stellar contrast, and the dust covering
factor. This does not imply that the preference for massive LRDs necessarily
disappears: the intrinsic AGN luminosity and black-hole mass must instead be
constrained from the extinction-corrected optical continuum and broad
H$\alpha$. This distinction is also relevant to the comparison with
\citet{Taylor2025_BHMF}, who retained uncorrected broad-H$\alpha$ luminosities
in their fiducial virial masses. They estimated that an internal-extinction
correction would increase the inferred masses by a median of 0.19 dex for the
full sample and 0.31 dex above $10^7\,M_\odot$, reaching 0.55 dex for the most
reddened source. Such corrections would move their BHMF toward the high-mass
tail predicted here. A self-consistent treatment of the stellar-dominated case
requires a two-component population model and is deferred to future work.}

\section{Summary}

We have presented a forward model in which LRDs are the dust-reddened,
high-inclination subset of the same \textcolor{black}{BLAGN population} that appears as
LBDs along cleaner sightlines. Starting from the observed \textcolor{black}{BLAGN UV}
LF, the model combines a super-Eddington, anisotropic continuum, an inclination-dependent obscuring cloud population, and the same photometric color selection used in current LRD samples.

The model reproduces the observed LRD LF using a modest luminosity-dependent
dust covering factor, whose average value is of order $\langle C_{\rm dust}\rangle\simeq 0.2$, together with characteristic cloud extinctions of a few magnitudes. These parameters are not independently fixed: the fit shows an elongated degeneracy in which lower cloud extinctions can be compensated by larger dust covering factors, while larger extinctions require a smaller intercepted fraction. The predicted apparent LRD/BLAGN fraction is
strongly luminosity dependent, rising from a few percent at the bright end to
$\sim20\%$ near $M_{1500}\simeq -19$.

A key prediction is that the apparent LRD/BLAGN ratio should be larger when
the population is binned by rest-frame optical magnitude than by UV magnitude.
This does not require a larger intrinsic obscured fraction at longer wavelengths, but follows from color-driven migration across observed magnitude bins: dust-reddened sources are strongly displaced in the UV while remaining comparatively bright in the optical.

Finally, the same best-fitting mock population predicts a strong segregation
in inferred black-hole mass between unobscured and dust-reddened sightlines.
This segregation is a consequence of the luminosity inversion rather than evidence for distinct parent populations. The resulting BHMF should therefore be interpreted as a model-dependent mass distribution for actively accreting \textcolor{black}{BLAGNs}, conditional on the assumed accretion-rate distribution and dust correction.

\label{lastpage}
\bibliographystyle{aa}
\bibliography{paper}

\begin{appendix}
\section{Clumpy dust obscuration model}

Following Paper I, we describe the dusty obscurer as a clumpy, equatorially concentrated cloud distribution by specifying the mean number of clouds intersected along a line of sight at inclination $i$ as
\begin{equation}
N_d(i)=N_{0,d}\,
\exp\!\left[-\frac{\cos^2 i}{2\sigma_d^2}\right],
\label{eq:Nlos_dust}
\end{equation}
(see, e.g., the formalism of \citealt{Nenkova2008}), where $\sigma_d$ controls the angular thickness of the cloud distribution about the equatorial plane. The corresponding escape probability for direct disk photons is
\begin{equation}
P_{{\rm esc},d}(i)=\exp\!\big[-N_d(i)\big].
\label{eq:Pdust}
\end{equation}
The normalization $N_{0,d}$ is fixed by requiring that the solid-angle-averaged probability of intercepting at least one dusty cloud equals the global dust covering factor,
\begin{equation}
C_{\rm dust}=\int_{0}^{\pi/2}\big[1-P_{{\rm esc},d}(i)\big]\sin i\,{\rm d}i.
\label{eq:Cdust}
\end{equation}
Here, $C_{\rm dust}$ is the solid-angle covering factor of the effective obscurer, i.e.\ the cloud population whose interception reddens the continuum sufficiently to produce a photometric LRD. The corresponding conditional inclination distribution of dust-intercepting sightlines is
\begin{equation}
p(i\,|\,d)=\frac{\big[1-P_{{\rm esc},d}(i)\big]\sin i}{C_{\rm dust}}.
\label{eq:pidust}
\end{equation}
Physically, a broader angular distribution for the dusty component than for the BLR is expected because the obscurer lies at larger radii, beyond the dust sublimation front, where radiation pressure on dust and/or disk winds can produce a larger scale height than in the dust-free BLR \citep[e.g.,][]{Elitzur2006,Nenkova2008,Elitzur2008,Wada2012}.

\end{appendix}
\end{document}